%% file: main.tex
\documentclass[conference]{IEEEtran}
\IEEEoverridecommandlockouts

\usepackage{cite,amsmath,amssymb,amsfonts,bbm,algorithmic,graphicx,textcomp,xcolor,tikz}
\usetikzlibrary{fit, positioning}
\def\BibTeX{{\rm B\kern-.05em{\sc i\kern-.025em b}\kern-.08em
    T\kern-.1667em\lower.7ex\hbox{E}\kern-.125emX}}

\newtheorem{theorem}{Theorem}
\newtheorem{lemma}{Lemma}

\newtheorem{definition} {Definition}
\newtheorem{remark}{Remark}

\newcommand{\relsm}[2]{\mathrel{\smash{\overset{\scriptscriptstyle(#1)}{#2}}}}
\newcommand{\relsmref}[2]{\mathrel{\smash{\overset{\scriptscriptstyle\eqref{#1}}{#2}}}}

\DeclareMathOperator{\EX}{\mathbb{E}}
\DeclareMathOperator{\T}{\mathcal{T}_{\epsilon}}

\begin{document}

\title{Secure Joint Source-Channel Coding of \\
Multimodal Semantic Sources}

\author{
\IEEEauthorblockN{Denis Kozlov, Mahtab Mirmohseni, and Rahim Tafazolli}
\IEEEauthorblockA{
Institute for Communication Systems, University of Surrey\\
Guildford GU2 7XH, U.K.\\
Email: \{d.kozlov, m.mirmohseni, r.tafazolli\}@surrey.ac.uk
}
}

\maketitle

\begin{abstract}
We study the problem of secure joint source-channel coding for multimodal semantic sources transmitted over noisy wiretap channels. The source model consists of $m$ modalities (e.g., image, audio, and sensor data), all represented as random variables. The encoder observes independent and identically distributed samples of an arbitrary non-empty subset of modalities, which are encoded, using wiretap coding with a help of a secret key shared with the legitimate receiver, and transmitted over a discrete memoryless wiretap channel. The legitimate receiver reconstructs all modalities. We extend the rate-distortion-perception characterisation to semantic unimodal and multimodal sources. We establish outer and inner bounds on the tradeoff between transmission rate, key rate, fidelity, and secrecy under per-modality distortion and perception constraints, and per-subset equivocation constraint. The bounds coincide for the less noisy wiretap channels. Our results show that the fundamental limit for secrecy consists of three operationally distinct components: the level of compression of each subset of modalities, the secret key rate, and the statistics of the wiretap channel.
\end{abstract}

\section{Introduction}
Understanding how multimodal semantic sources can be jointly represented and securely transmitted over wireless networks presents a set of information-theoretic challenges, such as characterisation of the fundamental limits of compression and secrecy. Some of these challenges are already addressed. For example, information-theoretic unimodal, i.e., with two correlated components, and multimodal semantic source models are introduced in \cite{liu2021indirect, liu2022}, together with the accompanying rate-distortion function characterisation. Since then, the \textit{unimodal} semantic source model is considered in a variety of settings such as source coding with side information \cite{guo2026srdf}, perceptual constraints \cite{chai2023}, joint source-channel coding (JSCC) \cite{stavrou2023jscc}, and game-theoretic formulations \cite{xiao2022}. However, some of the corresponding results for \textit{multimodal} sources are unexplored and constitute a current research gap.

In wireless communication, which is vulnerable to eavesdropping, secrecy is an important aspect of data transmission. In multimodal data transmission, some modalities can reveal sensitive information to an eavesdropper directly or via their correlation. Thus, we may require some modalities to have stricter secrecy requirements. Meanwhile, correlations across modalities mean that one modality can leak information about the others, requiring a per-subset secrecy rather than individual modality secrecy.

There are results for models with \textit{unimodal} semantic source and secrecy constraints, e.g., secure lossy source coding without a secret key \cite{yamamoto1983}, lossless secure source coding with a secret key \cite{yamamoto1994}, lossy secure source coding with a secret key \cite{guo2022}, and secure JSCC of unimodal semantic sources \cite{kozlov2026}. Extending these results to secure \textit{multimodal} models remains an open problem.

In this paper, we bridge this gap and consider secure JSCC of multimodal semantic sources. Compared to previous works with secrecy constraints \cite{guo2022,kozlov2026}, in this paper, the source is multimodal, the encoder has access to samples of an arbitrary non-empty subset of source components and the encoder accounts for the perception quality of each modality. To support our main coding theorems, we adapt the lossy source coding problem and the rate-distortion function presented in \cite{liu2022} to account for per-modality perception \cite{blau2019}, measured per letter \cite{chen2022rdpf}, and we consider the scenario where a transmitter may have direct access only to some modalities of the data, and we characterise the resulting multimodal rate-distortion-perception function (RDPF). The goal of this work is to characterise the tradeoff between transmission rate, fidelity, and secrecy, given constraints for \textit{each} modality within the model of secure JSCC of multimodal sources over wiretap channels. The main contribution of this work is a single-letter characterisation of this tradeoff for less noisy wiretap channels, together with outer and inner bounds for general wiretap channels.



\section{Problem Statement}
\label{sec:problem}
\subsection{Notation}
We denote random variables (r.v.s) with capital letters (e.g., $X$), their realisations with lowercase letters (e.g., $x$), and alphabets with calligraphic letters (e.g., $\mathcal{X}$). The probability mass function (pmf) of $X$ is $p_X(x)$ or $p(x)$ or $p_X$, $X \perp Y$ denotes independence, and $\mathcal{D}_{\mathcal{X}}$ is the probability simplex on $\mathcal{X}$. $\T^k(X)$ is the strong typical set. $\| p - q \|_{\mathrm{TV}} \doteq \frac{1}{2} \sum_{x} |p(x) - q(x)|$ is the total variation (TV) distance, $D(p\|q)$ is the Kullback-Leibler (KL) divergence. The binary entropy function is shown as $H_b(.)$
$S_i$ is the $i$-th modality of the source, reconstructions are denoted with a hat, e.g., $\hat{S}_i$, and tuples with bold font, e.g., $\bold{S}=(S_1,...,S_m)$. The superscripts $k$ and $n$ denote source and channel sequence lengths, respectively.
We denote: $[m] \doteq \{1, ..., m\}$, the set complement as $A^c \doteq [m] \setminus A$, all subsets without the empty set as $2^{[m]}_* \doteq \{A \subseteq [m] : A \neq \emptyset\}$, and $[x]^+ \doteq \max\{x, 0\}$. $S_{i,j}$ is the $j$-th sample of $S_i^k$, $\bold{S}_{A,j} \doteq \{S_{i,j}\}_{i \in A}$, $\bold{S}_j \doteq \bold{S}_{[m],j}$, and $\bold{S}^j_A \doteq (\bold{S}_{A,1}, ..., \bold{S}_{A,j})$. $\mathcal{B}(.)$ is the Bernoulli distribution, $\mathbbm{1}\{\cdot\}$ is the indicator function, $\oplus$ denotes XOR operator. Notation $\relsm{42}{=}$ means that the equality is based on equation (42).

\subsection{Problem Statement}
\label{subsec:problem}
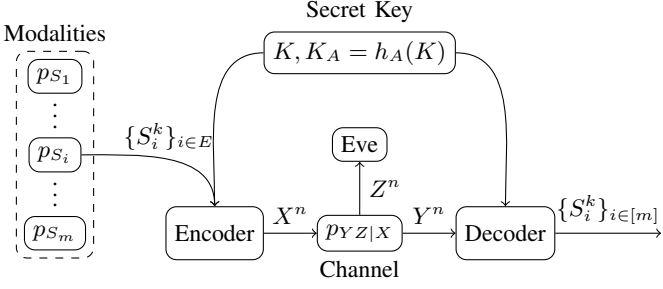
\begin{figure}\centering
    \input{model}
    \caption{System model.}
    \label{fig:model}
\end{figure}
We consider the problem (see Fig.~\ref{fig:model}) of secure transmission of a multimodal source consisting of $m$ correlated modalities $\bold{S} = (S_1,...,S_m)$, all modelled as discrete r.v.s with finite alphabets and the joint distribution $p_{\bold{S}} = p_{S_1,...,S_m}$. The encoder has direct access to the independent and identically distributed (i.i.d.) samples $\bold{S}^k_E$ of a non-empty subset of modalities $E \in 2_*^{[m]}$ and encodes them into the channel input $X^n$. The legitimate receiver and the eavesdropper observe $Y^n$ and $Z^n$, respectively, via the discrete memoryless wiretap channel $p_{YZ|X}$, and the legitimate receiver reconstructs all modalities $\hat{\bold{S}}^k = (\hat{S}^k_1, ..., \hat{S}^k_m)$. We assume that each reconstruction r.v. has finite alphabet. A secret key $K$, a uniform r.v., with limited rate $R_k$ is shared by the encoder and the legitimate decoder to enable secrecy via encryption. The first goal of the communication is to guarantee at the legitimate receiver average distortion $\EX d_i(S_i^k,\hat{S}_i^k)\le D_i$ for \textit{each} modality, i.e., $i \in [m]$, where $d_i(S_i^k,\hat{S}_i^k) \doteq \frac{1}{k} \sum_{j=1}^{k} d_i(S_{i,j}, \hat{S}_{i,j})$, and the $i$-th modality distortion measure is defined as $d_i : \mathcal{S}_i \times \hat{\mathcal{S}}_i \to [0,\infty)$. The second goal is to maintain the per-letter perception constraints $\phi_i(p_{S_i}, p_{\hat{S}_{i,j}}) \leq P_i$, $\forall j \in [k]$, where $\phi_i : \mathcal{D}_{\mathcal{S}_i} \times \mathcal{D}_{\hat{\mathcal{S}}_i} \to [0, \infty)$ is a perception metric with an assumption that it is convex and continuous in its second argument. The third goal is to maintain secrecy, measured by the equivocation, against the eavesdropper for \textit{each} modality and for every combination of modalities, i.e., $\frac{1}{k} H(\bold{S}^k_A|Z^n) \geq \Delta_A$, $\forall A \in 2^{[m]}_*$.
\begin{definition}[Code]\label{def:code}
An $(n,k)$ source-channel code consists of a stochastic encoder $f : \prod_{i \in E} \mathcal{S}^k_i \times \mathcal{K} \to \mathcal{X}^n$ and a stochastic decoder $g : \mathcal{Y}^n \times \mathcal{K} \to \prod_{i \in [m]} \hat{\mathcal{S}}^k_i$. \footnote{We assume that the randomness of the encoder and the decoder is private, i.e., independent of each other, of the source, of $K$, and of the channel noise.}
\end{definition}
\begin{definition}[Achievable Tuple]\label{def:ach}
Consider $\bold{D} = (D_1,...,D_m)$ to be a tuple of distortion limits, $\bold{P} = (P_1,...,P_m)$ to be a tuple of perception constraints, and equivocation requirements as $\boldsymbol{\Delta} = \{ \Delta_A \}_{A \in 2_*^{[m]}}$, a tuple $(r, R_k, \bold{D}, \bold{P}, \bold{\Delta})$ is achievable if for any $\epsilon > 0$ and large enough $k$ there exists an $(n,k)$ code with $\tfrac{n}{k} \leq r + \epsilon$ such that, for all $i \in [m]$ and $A \in 2_*^{[m]}$,
\begin{align}
    &\EX d_i(S^k_i,\hat{S}^k_i) \leq D_i + \epsilon, \ \ \max_{j \in [k]} \phi_i(p_{S_i}, p_{\hat{S}_{i,j}}) \leq P_i + \epsilon, \label{eq:ach-dp} \\
    &\tfrac{1}{k} H(\bold{S}_A^{k} | Z^{n}) \ge \Delta_A - \epsilon. \label{eq:ach-eqv}
\end{align}
\end{definition}
\begin{remark}
The joint equivocation constraints in addition to the individual ones may be necessary because the maximum of individual equivocations, in general, does not lead to the maximum of the joint equivocation. Consider the following example for $m = 2$. Let $S_1 \sim \mathcal{B}(0.5)$, $S_2 = S_1 \oplus B$, where $B \sim \mathcal{B}(0.5)$. In this case, $H(S_1) = H(S_2) = 1$ and $H(S_1,S_2) = 2$. Assume $Z = S_1 \oplus S_2$. Then, individual equivocations are at the maximum $H(S_1|Z) = 1$ and $H(S_2|Z) = 1$, while the joint $H(S_1,S_2|Z) = 1 \neq H(S_1|Z) + H(S_2|Z)$.
\end{remark}
\begin{definition}
\label{def:channel-capacity}
    $C \doteq \max_{p_X} I(X;Y)$ and $C_s \doteq \max_{p_{V,X}} (I(V;Y) - I(V;Z))$, where $V \to X \to (Y,Z)$ and $|\mathcal{V}| \leq |\mathcal{X}|$, are the capacity and the secrecy capacity of the wiretap channel, respectively. The wiretap channel is less noisy if $I(V;Y) \geq I(V;Z)$ for every $V \to X \to (Y,Z)$ \cite{elgamal2011book}.
\end{definition}

The goal of this work is to characterise the set of achievable tuples $(r, R_k, \bold{D}, \bold{P}, \bold{\Delta})$ in the asymptotic regime as $k \to \infty$.

\subsection{Multimodal Rate-Distortion-Perception Function}
\label{subsec:mrdpf}
We define the sub problem of lossy source coding for multimodal sources, where communication goals are the same as stated in Subsection~\ref{subsec:problem} but without channel and secrecy constraints, as follows. A tuple $(R, \bold{D}, \bold{P})$ is achievable if for every $\epsilon > 0$ and $k$ large enough there exist a stochastic encoder $f_s : \prod_{i \in E} \mathcal{S}^k_i \to \mathcal{W}$ and a stochastic decoder $g_s : \mathcal{W} \to \prod_{i \in [m]} \hat{\mathcal{S}}^k_i$, where the message rate is $\tfrac{1}{k} \log |\mathcal{W}| \leq R + \epsilon$, such that \eqref{eq:ach-dp} holds for all $i \in [m]$. The multimodal RDPF is defined as
\begin{equation}
    R(\bold{D}, \bold{P}) \doteq \inf \left\{ R : (R, \bold{D}, \bold{P}) \text{ is achievable} \right\}.
    \label{eq:rdpf-ach}
\end{equation}

\section{Main Results}\label{sec:results}
\subsection{Multimodal Rate-Distortion-Perception Function} 
\begin{lemma}[Multimodal RDPF]
\label{lemma:mrdpf}
For i.i.d. source $\bold{S}$ and the source coding problem presented in Subsection~\ref{subsec:mrdpf},
\begin{equation}\begin{aligned}
&R(\bold{D}, \bold{P}) = \min_{p_{ \hat{\bold{S}} | \bold{S}_E} } I(\bold{S}_E;\hat{\bold{S}}) \quad \text{s.t.} \\
&\EX d_i (S_i,\hat{S}_i) \leq D_i, \ \forall i \in E,  \\
&\EX \hat{d}_i (\bold{S}_E,\hat{S}_i) \leq D_i, \ \forall i \in E^c, \\
&\phi_i(p_{S_i}, p_{\hat{S}_i}) \leq P_i, \quad \forall i \in [m],
\label{eq:mrdpf}
\end{aligned}\end{equation}
where the modified distortion metric is
\begin{equation}
\hat{d}_i(\bold{s}_E,\hat{s}_i) = \textstyle\sum_{s_{i} \in \mathcal{S}_i} p(s_{i}|\bold{s}_{E}) d_i(s_{i},\hat{s}_{i}).
\label{eq:mdm}
\end{equation}
\end{lemma}
The proof of Lemma~\ref{lemma:mrdpf} is given in Subsection~\ref{subsec:lemma-proof}. We denote by $\mathcal{P}(\bold{D},\bold{P})$ the set of test channels $p_{\hat{\bold{S}}|\bold{S}_E}$ satisfying the constraints in \eqref{eq:mrdpf}.

\subsection{Outer and Inner Bounds}
\begin{theorem}[Outer Bound]
\label{theorem:outer}
If $(r, R_k, \bold{D}, \bold{P}, \bold{\Delta})$ is achievable, then there exists $p_{\hat{\bold{S}}|\bold{S}_E} \in \mathcal{P}(\bold{D},\bold{P})$ such that, for all $A \in 2_*^{[m]}$,
\begin{align}
&I(\bold{S}_E;\hat{\bold{S}}) \leq rC, \label{ineq:outer-rate} \\
&\Delta_A \leq H(\bold{S}_A) - \bigl[ I(\bold{S}_A;\hat{\bold{S}}) - R_k - rC_s \bigr]^+,\label{ineq:outer-secrecy}
\end{align}
where $C$ and $C_s$ are defined in Definition~\ref{def:channel-capacity}.
\begin{remark}
    If the wiretap channel is less noisy, then $C$ and $C_s$ can be replaced with $I(X;Y)$ and $I(X;Y) - I(X;Z)$, respectively, for some $p_X$.
\end{remark}

\end{theorem}
\begin{theorem}[Inner Bound]
\label{theorem:inner}
A tuple $(r, R_k, \bold{D}, \bold{P}, \bold{\Delta})$ is achievable if there exist $p_{\hat{\bold{S}}|\bold{S}_E} \in \mathcal{P}(\bold{D},\bold{P})$ and $p_{V,X}$ with $V \to X \to (Y,Z)$ such that, for all $A \in 2_*^{[m]}$,
\begin{align}
&I(\bold{S}_E;\hat{\bold{S}}) \leq r I(V;Y), \label{ineq:inner-rate} \\
&\Delta_A \leq H(\bold{S}_A) - \bigl[ I(\bold{S}_A;\hat{\bold{S}}) - R_k - r c_V \bigr]^+, \label{ineq:inner-secrecy}
\end{align}
where $c_V \doteq [I(V;Y) - I(V;Z)]^+$.
\end{theorem}
The proofs of Theorems~\ref{theorem:outer} and~\ref{theorem:inner} are presented in Subsections~\ref{subsec:outer-proof} and~\ref{subsec:inner-proof}, respectively. 
\begin{remark}
The inner and outer bounds differ only in the channel capacity terms and they coincide whenever $p_{V,X}$ attains both $C$ and $C_s$, i.e., $I(V;Y) = C$ and $c_V = C_s$. For example, for a noiseless wiretap channel $Y = Z = X$, where $C_s = 0$, or a binary symmetric wiretap channel with crossover probabilities $p_Y, p_Z \leq \frac{1}{2}$, where a uniform $V = X$ gives $I(V;Y) = C$ and $c_V = [H_b(p_Z) - H_b(p_Y)]^+ = C_s$. Moreover, for less noisy channels, $V = X$ in Theorem~\ref{theorem:inner} gives $c_V = I(X;Y) - I(X;Z)$ and matches Theorem~\ref{theorem:outer}.
\end{remark}

Theorems~\ref{theorem:outer} and~\ref{theorem:inner} show that the secrecy limit of every subset $A$ consists of three operationally distinct components: i) the compression leakage $I(\bold{S}_A;\hat{\bold{S}})$, ii) the secret key rate $R_k$, and iii) the wiretap coding, whose contribution depends on the statistics of the wiretap channel. A single test channel determines both the rate and the compression leakages of all subsets. In general, different test channels from $\mathcal{P}(\bold{D},\bold{P})$ distribute compression leakages differently which leads to a tradeoff between the transmission rate and the secrecy of different subsets of modalities.


\section{Proofs}
\label{sec:proofs}
\subsection{Proof of Lemma~\ref{lemma:mrdpf}}
\label{subsec:lemma-proof}
Let $R^{\mathrm{I}}(\bold{D}, \bold{P})$ denote the right-hand side of \eqref{eq:mrdpf}. The set of test channels $p_{\hat{\bold{S}}|\bold{S}_E}$ is compact because of finite source and reconstruction alphabets, and $I(\bold{S}_E;\hat{\bold{S}})$, the expected distortions, and $\phi_i(p_{S_i},p_{\hat{S}_i})$ are continuous functions of $p_{\hat{\bold{S}}|\bold{S}_E}$. Hence, the set of test channels $p_{\hat{\bold{S}}|\bold{S}_E}$ satisfying the constraints in \eqref{eq:mrdpf} is compact. Also, $\bigcap_{\epsilon > 0} \mathcal{P}(\bold{D} + \epsilon,\bold{P} + \epsilon) = \mathcal{P}(\bold{D},\bold{P})$, and we have:
\begin{equation}
    \lim_{\epsilon \to 0^+} R^{\mathrm{I}}(\bold{D} + \epsilon, \bold{P} + \epsilon) = R^{\mathrm{I}}(\bold{D}, \bold{P}).
    \label{eq:right-cont}
\end{equation}

\subsubsection{Converse}
Let $(R,\bold{D},\bold{P})$ be achievable, fix $\epsilon > 0$ and fix the encoder and decoder $(f_s, g_s)$. Let $J$ be a r.v. uniform on $[k]$ and independent of $(\bold{S}^k, W, \hat{\bold{S}}^k)$. We first bound the message rate.
\begin{align}
    &k(R + \epsilon) \relsm{a}{\geq} H(W) \relsm{b}{\geq} I(\bold{S}_E^k; \hat{\bold{S}}^k) \relsm{c}{=} \textstyle\sum_{j=1}^{k} I(\bold{S}_{E,j}; \hat{\bold{S}}^k, \bold{S}_E^{j-1}) \nonumber \\
    &{\geq} {\textstyle\sum_{j=1}^{k}} I(\bold{S}_{E,j}; \hat{\bold{S}}_j) {=} k I(\bold{S}_{E,J}; \hat{\bold{S}}_J | J) {\relsm{d}{\geq}} k I(\bold{S}_{E,J}; \hat{\bold{S}}_J),
    \label{eq:rdpf-converse-chain}
\end{align}
where (a) is due to $\tfrac{1}{k} H(W) \leq \tfrac{1}{k} \log |\mathcal{W}| \leq R + \epsilon$, (b) is due to data processing inequality (DPI), (c) is due to the chain rule and $I(\bold{S}_{E,j}; \bold{S}_E^{j-1}) = 0$ because the source is i.i.d, (d) is because $J \perp \bold{S}_{E,J}$.

We next analyse the average distortion. For $i \in [m]$ we have
\begin{align}
    &\EX d_i(S_{i,J}, \hat{S}_{i,J}) \relsm{a}{=} \textstyle\sum_{j=1}^{k} \Pr(J {=} j) \EX \left[ d_i(S_{i,J}, \hat{S}_{i,J})|J=j \right] \nonumber \\
    &\relsm{b}{=} \tfrac{1}{k} \textstyle\sum_{j=1}^{k} \EX d_i(S_{i,j}, \hat{S}_{i,j}) \relsm{c}{=} \EX d_i(S_i^k, \hat{S}_i^k) \relsmref{eq:ach-dp}{\leq} D_i + \epsilon,
    \label{eq:rdpf-converse-dist-E}
\end{align}
where (a) is due to the law of total expectation, (b) is because $J$ is uniform and $J \perp \bold{S}^k_{i}, \hat{\bold{S}}^k_{i}$, (c) is due to linearity of expectation and the definition of the block distortion $d_i(S^k, \hat{S}^k)$. For $i \in E^c$ we also have
\begin{align}
    &\EX d_i(S_{i,J}, \hat{S}_{i,J}) 
    \relsm{a}{=} \tfrac{1}{k} \textstyle\sum_{j=1}^{k} \EX\bigl[ \EX[d_i(S_{i,j}, \hat{S}_{i,j}) | \bold{S}_{E,j}, \hat{S}_{i,j}] \bigr] \nonumber \\
    &\relsm{b}{=} \tfrac{1}{k} \textstyle\sum_{j=1}^{k} \EX\Bigl[ \textstyle\sum_{s_i \in \mathcal{S}_i} p(s_i | \bold{S}_{E,j}) d_i(s_i, \hat{S}_{i,j}) \Bigr] \nonumber \\
    &\overset{\eqref{eq:mdm}}{=} \tfrac{1}{k} \textstyle\sum_{j=1}^{k} \EX \hat{d}_i(\bold{S}_{E,j}, \hat{S}_{i,j}) = \EX \hat{d}_i(\bold{S}_{E,J}, \hat{S}_{i,J}),
    \label{eq:rdpf-converse-dist-Ec-derivation}
\end{align}
where (a) is the tower property of conditional expectation; (b) is due to $S_{i,j} \to \bold{S}_{E,j} \to \hat{S}_{i,j}, \ \forall j \in [k]$. So for the same reasons as in \eqref{eq:rdpf-converse-dist-E}, we then have $\EX \hat{d}_i(\bold{S}_{E,J}, \hat{S}_{i,J}) \leq D_i + \epsilon$.

The perception analysis for $i \in [m]$ is as follows. Given that $J$ is uniform and $J \perp \hat{\bold{S}}^k$, we have $p_{\hat{S}_{i,J}}(\hat{s}) = \textstyle\sum_{j=1}^{k} \Pr(J = j) \Pr(\hat{S}_{i,j} = \hat{s} | J = j) = \tfrac{1}{k} \textstyle\sum_{j=1}^{k} p_{\hat{S}_{i,j}}(\hat{s})$. Hence,
\begin{align}
    &\phi_i (p_{S_i}, p_{\hat{S}_{i,J}})
    \relsm{a}{\leq} \tfrac{1}{k} \textstyle\sum_{j=1}^{k} \phi_i (p_{S_i}, p_{\hat{S}_{i,j}} )
    \overset{\eqref{eq:ach-dp}}{\leq} P_i + \epsilon,
    \label{eq:rdpf-converse-perc}
\end{align}
where (a) is due to $p_{\hat{S}_{i,J}}(\hat{s}) = \tfrac{1}{k} \textstyle\sum_{j=1}^{k} p_{\hat{S}_{i,j}}(\hat{s})$, the convexity of $\phi_i(p_{S_i},\cdot)$, and Jensen's inequality.

Since $\bold{S}_{E,J} \sim p_{\bold{S}_E}$ the test channel $p_{\hat{\bold{S}}_J|\bold{S}_{E,J}}$ is in $\mathcal{P}(\bold{D} + \epsilon,\bold{P} + \epsilon)$ because of \eqref{eq:rdpf-converse-dist-E}-\eqref{eq:rdpf-converse-perc}. By \eqref{eq:rdpf-converse-chain}, $R + \epsilon \geq I(\bold{S}_{E,J};\hat{\bold{S}}_J) \geq R^{\mathrm{I}}(\bold{D} + \epsilon,\bold{P} + \epsilon)$. Letting $\epsilon \to 0$ and using \eqref{eq:right-cont} gives $R \geq R^{\mathrm{I}}(\bold{D},\bold{P})$ for every achievable $R$, i.e., $R(\bold{D},\bold{P}) \geq R^{\mathrm{I}}(\bold{D},\bold{P})$.

\subsubsection{Achievability}
Fix the test channel $p_{\hat{\bold{S}} | \bold{S}_E}$ attaining $R^{\mathrm{I}}(\bold{D},\bold{P}) < \infty$, $\epsilon > 0$, and $R > R^{\mathrm{I}}(\bold{D},\bold{P})$. Define $N = 2^{kR}$. Generate $N$ codewords $\{ \hat{\bold{s}}^k(l) \}$, $l \in [N]$, with i.i.d. entries from $ p_{\hat{\bold{S}}}$. We use the likelihood encoder \cite{song2016}, which, given $\bold{s}^k_E$, draws the index $l$ with probability
\begin{equation}
    \ell(l|\bold{s}^k_E) = \frac{\prod_{j=1}^{k} p_{\bold{S}_E|\hat{\bold{S}}}(\bold{s}_{E,j}|\hat{\bold{s}}_j(l))}{\sum_{l'=1}^{N} \prod_{j=1}^{k} p_{\bold{S}_E|\hat{\bold{S}}}(\bold{s}_{E,j}|\hat{\bold{s}}_j(l'))}
    \label{eq:likelihood}
\end{equation}
Assume that the probability is uniform if the denominator of \eqref{eq:likelihood} is zero. The decoder outputs $\hat{\bold{S}}^k = \hat{\bold{s}}^k(L)$, where $L$ represents random index $l$. For a fixed codebook and for i.i.d source, this coding scheme induces
\begin{equation}
    \mathsf{P}(\bold{s}^k, l) = \left( \textstyle\prod_{j \in [k]} p_{\bold{S}}(\bold{s}_j) \right) \ell(l|\bold{s}^k_E).
    \label{eq:mrdpf-P}
\end{equation}
We consider the idealised distribution: $L$ is uniform, and the source is generated from the selected codeword through the backward channel $ p_{\bold{S}|\hat{\bold{S}}}$, i.e.,
\begin{equation}
    \mathsf{Q}(\bold{s}^k, l) = \tfrac{1}{N} \textstyle\prod_{j=1}^{k} p_{\bold{S}|\hat{\bold{S}}}(\bold{s}_j|\hat{\bold{s}}_j(l)).
    \label{eq:mrdpf-Q}
\end{equation}
We now analyse $T \doteq \| \mathsf{P} - \mathsf{Q} \|_{\mathrm{TV}}$. Note that
\begin{equation}
    p_{\bold{S}|\hat{\bold{S}}} {=} p_{\bold{S}_{E^c},\bold{S}_{E}|\hat{\bold{S}}}
    \relsm{a}{=} \tfrac{ p_{\bold{S}_{E}} p_{\bold{S}_{E^c}|\bold{S}_{E}} p_{\hat{\bold{S}} | \bold{S}_{E}} } { p_{\hat{\bold{S}}} } = p_{\bold{S}_{E} | \hat{\bold{S}}} p_{ \bold{S}_{E^c} | \bold{S}_{E} }
    \label{eq:backward-channel-factorisation}
\end{equation}
where (a) is due to $\bold{S}_{E^c} \to \bold{S}_{E} \to \hat{\bold{S}}$. Also note
\begin{align*}
    &\mathsf{Q}(\bold{s}^k) = \textstyle\sum_{l=1}^{N} \mathsf{Q}(\bold{s}^k,l) \overset{\eqref{eq:backward-channel-factorisation}}{=} \mathsf{Q}(\bold{s}^k_E) \textstyle\prod_{j=1}^{k} p_{\bold{S}_{E^c}|\bold{S}_E}(\bold{s}_{E^c,j}|\bold{s}_{E,j}),
\end{align*}
where $\mathsf{Q}(\bold{s}^k_E) = \tfrac{1}{N} \textstyle\sum_{l=1}^{N} \prod_{j=1}^{k} p_{\bold{S}_E|\hat{\bold{S}}}(\bold{s}_{E,j}|\hat{\bold{s}}_j(l))$. We have:
\begin{align}
&\textstyle\sum_{l=1}^{N} \bigl|\mathsf{P}(\bold{s}^k, l) - \mathsf{Q}(\bold{s}^k, l)\bigr| \\
&\relsm{a}{=} \textstyle\sum_{l=1}^{N} \bigl|\left( \textstyle\prod_{j=1}^{k} p_{\bold{S}}(\bold{s}_j) \right) \ell(l|\bold{s}^k_E) - \mathsf{Q}(\bold{s}^k) \ell(l|\bold{s}^k_E)\bigr|   \\
&\relsm{b}{=} \bigl|\textstyle\prod_{j=1}^{k}  p_{\bold{S}}(\bold{s}_j) - \mathsf{Q}(\bold{s}^k)\bigr|
\label{eq:partial-tv-sum}
\end{align}
where (a) is due to \eqref{eq:mrdpf-P}, \eqref{eq:mrdpf-Q} and $\mathsf{Q}(l|\bold{s}^k) = \tfrac{\mathsf{Q}(\bold{s}^k,l)}{\mathsf{Q}(\bold{s}^k)} = \ell(l|\bold{s}^k_E)$, (b) is because of the common factor $\textstyle\sum_{l=1}^{N} \ell(l|\bold{s}^k_E) = 1$. Hence for $T$ we have:
\begin{align}
&T = \tfrac{1}{2} \textstyle\sum_{\bold{s}^k} \textstyle\sum_{l=1}^{N} \bigl|\mathsf{P}(\bold{s}^k, l) {-} \mathsf{Q}(\bold{s}^k, l)\bigr| \\
&\overset{\eqref{eq:partial-tv-sum}}{=} \tfrac{1}{2} \textstyle\sum_{\bold{s}^k} \bigl|\textstyle\prod_{j} p_{\bold{S}}(\bold{s}_j) 
- \mathsf{Q}(\bold{s}^k)\bigr| \\
&= \tfrac{1}{2} \textstyle\sum_{\bold{s}^k_E} \bigl|\textstyle\prod_{j} p_{\bold{S}_E}(\bold{s}_{E,j}) - \mathsf{Q}(\bold{s}^k_E)\bigr| \textstyle\sum_{\bold{s}^k_{E^c}} C(\bold{s}_{E^c}^k) \label{eq:common-factor} \\
&= \| \textstyle\prod_{j} p_{\bold{S}_E}(\bold{s}_{E,j}) - \mathsf{Q}_{\bold{S}^k_E} \|_{\mathrm{TV}},
\label{eq:lemma-inner-tv-reduction}
\end{align}
where $C(\bold{s}_{E^c}^k) = \prod_{j=1}^{k} p_{\bold{S}_{E^c}|\bold{S}_E}(\bold{s}_{E^c,j}|\bold{s}_{E,j})$ is the common factor with a sum of $1$. By the soft covering lemma \cite[Lemma 1]{song2016}, given $R > I(\bold{S}_E;\hat{\bold{S}})$, and Markov inequality we have $\EX T \to 0$ and $\Pr(T > \epsilon) \leq \tfrac{\EX T}{\epsilon}$ where the expectation is over the random codebook.

We now analyse average distortion for the proposed coding scheme under the idealised distribution $\mathsf{Q}$.
\begin{align}
&\EX_{\mathsf{Q}} d_i(S_i^k, \hat{S}_i^k) \relsm{a}{=} \tfrac{1}{N} \textstyle\sum_{l=1}^{N} \EX_{\mathsf{Q}}\bigl[ d_i(S_i^k, \hat{S}_i^k) \big| L = l \bigr] \\
&\relsm{b}{=} \tfrac{1}{N} \textstyle\sum_{l=1}^{N} \tfrac{1}{k} \sum_{j=1}^{k} \EX_{\mathsf{Q}}\bigl[ d_i(S_{i,j}, \hat{s}_{i,j}(l)) \big| L = l \bigr] \label{eq:mrdpf-dist-Q-prev} \\
&\relsm{c}{=} \tfrac{1}{kN} \textstyle\sum_{l=1}^{N} \sum_{j=1}^{k} d^*_i(\hat{\bold{s}}_j(l)),
\label{eq:mrdpf-dist-Q}
\end{align}
where (a) is due to the law of total expectation and $L$ being uniform under $\mathsf{Q}$; (b) is due to the definition of $d_i(S_i^k, \hat{S}_i^k)$, the linearity of expectation, and $\hat{\bold{S}}^k = \hat{\bold{s}}^k(l)$ given $L = l$; and (c) holds because under $\mathsf{Q}$, $S_{i,j} \sim p_{S_i|\hat{\bold{S}}}(s_i|\hat{\bold{s}}_j(l))$, and we define $d^*_i(\hat{\bold{s}}) = \sum_{s_i} p_{S_i|\hat{\bold{S}}}(s_i|\hat{\bold{s}}) d_i(s_i,\hat{s}_i)$. Over the random codebook, the $kN$ terms $d^*_i(\hat{\bold{s}}_j(l))$ in \eqref{eq:mrdpf-dist-Q} are i.i.d., since the codebook entries are i.i.d. $\sim p_{\hat{\bold{S}}}$. They take values in $[0, d_{\max}]$, where $d_{\max} = \max_{i,s,\hat{s}} d_i(s,\hat{s})$. Note that since $p_{\hat{\bold{S}}|\bold{S}_E} \in \mathcal{P}(\bold{D},\bold{P})$, we have $\EX d^*_i(\hat{\bold{S}}) = \EX d_i(S_i,\hat{S}_i) \leq D_i$, and for $i \in E^c$ also, $\EX d_i(S_i,\hat{S}_i) = \EX \hat{d}_i(\bold{S}_E,\hat{S}_i)$ due to $S_i \to \bold{S}_E \to \hat{S}_i$. We have
\begin{align}
&\Pr\bigl(\EX_{\mathsf{Q}} d_i(S_i^k,\hat{S}_i^k) > D_i + \epsilon\bigr) \label{eq:event1} \\
&\relsm{a}{\leq} \Pr\bigl(\bigl|\EX_{\mathsf{Q}} d_i(S_i^k,\hat{S}_i^k) - \EX d^*_i(\hat{\bold{S}})\bigr| \geq \epsilon\bigr)
\relsm{b}{\leq} \frac{d_{\max}^2}{4kN\epsilon^2},
\label{eq:event2} 
\end{align}
where (a) is because $\EX d^*_i(\hat{\bold{S}}) \leq D_i$ so the first event in \eqref{eq:event1} in included in the second \eqref{eq:event1}, and (b) is due to Chebyshev's inequality. With $k \to \infty$ we have $\EX_{\mathsf{Q}} d_i(S_i^k,\hat{S}_i^k) \leq D_i + \epsilon$ with high probability (w.h.p).

The perception analysis is as follows. Under $\mathsf{Q}$, $L$ is uniform and $\hat{\bold{S}}^k = \hat{\bold{s}}^k(L)$, so, for every $j \in [k]$,
$\mathsf{Q}_{\hat{S}_{i,j}}(\hat{s}) = \tfrac{1}{N} \textstyle\sum_{l=1}^{N} \mathbbm{1}\{\hat{s}_{i,j}(l) = \hat{s}\}$. The indicators are independent, take values in $[0,1]$, and have mean $p_{\hat{S}_i}(\hat{s})$. Hence, by Hoeffding's inequality, $\Pr\bigl(|\mathsf{Q}_{\hat{S}_{i,j}}(\hat{s}) - p_{\hat{S}_i}(\hat{s})| > \tfrac{2\epsilon}{|\hat{\mathcal{S}}_i|}\bigr) \leq 2e^{-8N\epsilon^2/|\hat{\mathcal{S}}_i|^2}$.
If no deviation outside $\tfrac{2\epsilon}{|\hat{\mathcal{S}}_i|}$ occurs for any $\hat{s}$, then
\begin{align*}
\|\mathsf{Q}_{\hat{S}_{i,j}} {-} p_{\hat{S}_i}\|_{\mathrm{TV}} &= \tfrac{1}{2} \textstyle\sum_{\hat{s}} |\mathsf{Q}_{\hat{S}_{i,j}}(\hat{s}) - p_{\hat{S}_i}(\hat{s})| \leq \tfrac{1}{2} |\hat{\mathcal{S}}_i| \tfrac{2\epsilon}{|\hat{\mathcal{S}}_i|} = \epsilon.
\end{align*}
Hence, by the union bound over $j \in [k]$ and $\hat{s} \in \hat{\mathcal{S}}_i$,
\begin{equation*}
    \Pr\Bigl( \max_{j \in [k]} \| \mathsf{Q}_{\hat{S}_{i,j}} - p_{\hat{S}_i} \|_{\mathrm{TV}} > \epsilon \Bigr) \leq 2k |\hat{\mathcal{S}}_i| e^{-8N\epsilon^2/|\hat{\mathcal{S}}_i|^2},
\end{equation*}
which vanishes, since $N$ grows exponentially in $k$.

One can fix a codebook for which all relevant in this proof events hold. Due to $|\EX_{\mathsf{P}} f - \EX_{\mathsf{Q}} f| \leq f_{\max} T$ for every function $f$ of $(\bold{S}^k, L)$ with values in $[0, f_{\max}]$, we have $\EX_{\mathsf{P}} d_i(S_i^k, \hat{S}_i^k) \leq D_i + (1 + d_{\max})\epsilon$ and, with indicator functions $f$, $\| \mathsf{P}_{\hat{S}_{i,j}} - p_{\hat{S}_i} \|_{\mathrm{TV}} \leq T + \epsilon \leq 2\epsilon$ for all $j \in [k]$. Since the perception metric $\phi_i(p_{S_i},\cdot)$ is continuous, and $\phi_i(p_{S_i}, p_{\hat{S}_i}) \leq P_i$, $\phi_i(p_{S_i}, \mathsf{P}_{\hat{S}_{i,j}}) \leq P_i + \delta$ for all $j \in [k]$, where $\delta \to 0$. Since $\epsilon > 0$ is arbitrary, every $R > R^{\mathrm{I}}(\bold{D},\bold{P})$ is achievable, i.e., $R(\bold{D},\bold{P}) \leq R^{\mathrm{I}}(\bold{D},\bold{P})$.

\subsection{Proof of Theorem~\ref{theorem:outer}}
\label{subsec:outer-proof}
Fix $\epsilon > 0$ and a code satisfying Definition~\ref{def:ach}. Let $J$ be uniform r.v. on $[k]$, and independent of all other system r.v.s. From the problem statement, the encoder generates $X^n$ from $(\bold{S}^k_E, K)$ and the decoder generates $\hat{\bold{S}}^k$ from $(Y^n, K)$, where $K$ and the channel noise are independent of the source. Hence, $\bold{S}^k \to \bold{S}^k_E \to \hat{\bold{S}}^k$, and, as in the proof of Lemma~\ref{lemma:mrdpf}, $\bold{S}_{E^c,J} \to \bold{S}_{E,J} \to \hat{\bold{S}}_J$, $(\bold{S}_J, \hat{\bold{S}}_J) \sim p_{\bold{S}} p_{\hat{\bold{S}}_J|\bold{S}_{E,J}}$. The distortion and perception analysis steps then follow from \eqref{eq:rdpf-converse-dist-E}-\eqref{eq:rdpf-converse-perc} as they use only the Markov chain and \eqref{eq:ach-dp}, so we have $p_{\hat{\bold{S}}_J|\bold{S}_{E,J}} \in \mathcal{P}(\bold{D}+\epsilon,\bold{P}+\epsilon)$. We continue with the proof \eqref{ineq:outer-rate}:
\begin{align}
&k I(\bold{S}_{E,J}; \hat{\bold{S}}_J) \relsmref{eq:rdpf-converse-chain}{\leq} I(\bold{S}^k_E; \hat{\bold{S}}^k) \relsm{a}{\leq} I(X^n; Y^n) \\ 
&\relsm{b}{\leq} n I(X;Y) \leq nC, \label{eq:outer-rate-chain}
\end{align}
where (a) is due to DPI and independence properties of $K$, (b) holds since the channel is memoryless. The proof of \eqref{ineq:outer-secrecy} as follows. For every $A \in 2^{[m]}_*$,
\begin{align}
&k (H(\bold{S}_A) {-} \Delta_A {+} \epsilon) \relsm{a}{\geq} I(\bold{S}^k_A; Z^n) \\
&= I(\bold{S}^k_A; Y^n) - ( I(\bold{S}^k_A; Y^n) - I(\bold{S}^k_A; Z^n) ) \\
&\relsm{b}{\geq} I(\bold{S}^k_A; Y^n, K) - H(K) - n C_s \\
&\relsm{c}{\geq} I(\bold{S}^k_A; \hat{\bold{S}}^k) - k R_k - n C_s \\
&\relsmref{eq:rdpf-converse-chain}{\geq} k I(\bold{S}_{A,J}; \hat{\bold{S}}_J) - k R_k - n C_s \label{eq:outer-secrecy-chain}
\end{align}
where (a) is due to \eqref{eq:ach-eqv} and the i.i.d source; (b) for the same reasons as in \cite[(38)-(43)]{kozlov2026}, (c) is due to DPI and properties of $K$, i.e., the uniformity and independence.

To finish the proof, we divide \eqref{eq:outer-rate-chain} and \eqref{eq:outer-secrecy-chain} by $k$ and use $n \leq k(r+\epsilon)$ and the trivial bound $\Delta_A - \epsilon \leq \frac{1}{k} H(\bold{S}^k_A|Z^n) \leq H(\bold{S}_A)$. As $\epsilon \to 0$, by the compactness property, a subsequence of test channels $p_{\hat{\bold{S}}_J|\bold{S}_{E,J}} \in \mathcal{P}(\bold{D}+\epsilon,\bold{P}+\epsilon)$ converges to some $p_{\hat{\bold{S}}|\bold{S}_E}$, which is in $\mathcal{P}(\bold{D},\bold{P})$, as in \eqref{eq:right-cont}, and satisfies \eqref{ineq:outer-rate}-\eqref{ineq:outer-secrecy} for all subsets $A$.

\subsection{Proof of Theorem~\ref{theorem:inner}}
\label{subsec:inner-proof}
Fix $p_{\hat{\bold{S}}|\bold{S}_E}$ and $p_{V,X}$ as in Theorem~\ref{theorem:inner}, and fix $\epsilon > 0$. Fix $r' \in (r, r + \epsilon)$, $n = \lceil r'k \rceil$, and $\delta > 0$ such that the rate is $R \doteq r'I(V;Y) - \delta > I(\bold{S}_E;\hat{\bold{S}})$. We generate a source codebook $\{\hat{\bold{s}}^k(\kappa,l)\}$, $(\kappa,l) \in \mathcal{K} \times [N]$, where $N \doteq 2^{kR}$, with i.i.d.\ entries distributed as $p_{\hat{\bold{S}}}$, where $\{\hat{\bold{s}}^k(\kappa,l)\}_{l}$ is the sub-codebook of $\kappa$. Independently of the source codebook we generate a channel codebook $\{v^n(l)\}$, $l \in [N]$, with i.i.d. entries distributed as $p_V$. We use the likelihood encoder, which, given the encoder input $\bold{s}^k_E$ and the secret key $K = \kappa$, draws the index $l$ with probability $\ell(l|\bold{s}^k_E,\kappa)$, given by \eqref{eq:likelihood} for the sub-codebook of $\kappa$, and transmits $X^n \sim \prod_{t=1}^{n} p_{X|V}(x_t|v_t(l))$. The decoder finds the unique $\hat{l}$ such that $(v^n(\hat{l}), Y^n) \in \T^n(V,Y)$ ($\hat{l} = 1$ if there is none or more than one) and outputs $\hat{\bold{s}}^k(\kappa,\hat{l})$. For a fixed codebook, a uniform secret key, the i.i.d. source, and the memoryless wiretap channel, the scheme induces
\begin{align}
&\mathsf{P}(\bold{s}^k, \kappa, l, x^n, y^n, z^n) = 2^{-kR_k} \left( \textstyle\prod_j p_{\bold{S}}(\bold{s}^k_j) \right) \ell(l|\bold{s}^k_E,\kappa) \nonumber \\
&\quad \cdot \textstyle\prod_{t=1}^{n} p_{X|V}(x_t|v_t(l)) p_{YZ|X}(y_t,z_t|x_t).
\label{eq:inner-P}
\end{align}
Consider the idealised distribution, under which $(K,L)$ is uniform and the source is generated from the selected codeword through $p_{\bold{S}|\hat{\bold{S}}}$:
\begin{align}
&\mathsf{Q}(\bold{s}^k, \kappa, l, x^n, y^n, z^n) = 2^{-k(R_k + R)} \textstyle\prod_{j=1}^{k} p_{\bold{S}|\hat{\bold{S}}}(\bold{s}_j|\hat{\bold{s}}_j(\kappa,l)) \nonumber \\
&\quad \cdot \textstyle\prod_{t=1}^{n} p_{X|V}(x_t|v_t(l)) p_{YZ|X}(y_t,z_t|x_t).
\label{eq:inner-Q}
\end{align}
 
Given $K = \kappa$, the pmfs of $(\bold{S}^k, L)$ under \eqref{eq:inner-P} and \eqref{eq:inner-Q} are \eqref{eq:mrdpf-P} and \eqref{eq:mrdpf-Q} for the sub-codebook of $\kappa$. Hence, for $T_K \doteq \|\mathsf{P} - \mathsf{Q}\|_{\mathrm{TV}}$, we have,
\begin{align}
T_K &\relsm{a}{=} 2^{-kR_k} \textstyle\sum_{\kappa} \tfrac{1}{2} \sum_{\bold{s}^k, l} \bigl|\mathsf{P}(\bold{s}^k,l|\kappa) - \mathsf{Q}(\bold{s}^k,l|\kappa)\bigr| \nonumber \\
&\relsm{b}{=} 2^{-kR_k} \textstyle\sum_{\kappa} \| \textstyle\prod_j p_{\bold{S}_E} - \mathsf{Q}_{\bold{S}^k_E|K=\kappa} \|_{\mathrm{TV}}, \label{eq:inner-approx}
\end{align}
where (a) holds since $\mathsf{P}(\kappa) = \mathsf{Q}(\kappa) = 2^{-kR_k}$, and the channel factor in \eqref{eq:inner-P}-\eqref{eq:inner-Q} is common, depends only on $l$, and sums to one over $(x^n,y^n,z^n)$; and (b) is for the same reasons as in \eqref{eq:lemma-inner-tv-reduction} for the sub-codebook of $\kappa$. Each sub-codebook is an i.i.d. codebook of rate $R > I(\bold{S}_E;\hat{\bold{S}})$, so, by the soft-covering lemma \cite[Lemma~1]{song2016}, every term in \eqref{eq:inner-approx} has vanishing expectation. Hence, $\EX T_K \to 0$, and, by Markov's inequality, $T_K \leq \epsilon$ w.h.p.
 
Let $\mathsf{Q}'$ be $\mathsf{Q}$ with the decoder output replaced by $\hat{\bold{s}}^k(\kappa, l)$ instead of $\hat{\bold{s}}^k(\kappa, \hat{l})$. Under $\mathsf{Q}$, $L$ is uniform, and $Y^n$ is the output of the memoryless channel $p_{Y|V}$ with input $v^n(L)$. Since $kR/n \leq I(V;Y) - \delta/(r'+1)$, with the help of the packing lemma \cite[Lemma~3.1]{elgamal2011book}, we have $\mathsf{Q}(\hat{L} \neq L) \leq \epsilon$ w.h.p. As $\mathsf{Q}$ and $\mathsf{Q}'$ differ only in the decoder output on $\{\hat{L} \neq L\}$, using the triangle inequality we have $\|\mathsf{P} - \mathsf{Q}'\|_{\mathrm{TV}} \leq T_K + \mathsf{Q}(\hat{L} \neq L) \leq 2\epsilon$ w.h.p.
 
We now analyse the distortion and perception for the proposed scheme. Under $\mathsf{Q}'$, $(K,L)$ is uniform over all $N' = 2^{k(R_k + R)}$ codewords, $\bold{S}^k$ is generated from $\hat{\bold{s}}^k(K,L)$ through $p_{\bold{S}|\hat{\bold{S}}}$, and the decoder outputs $\hat{\bold{s}}^k(K,L)$, i.e., $\mathsf{Q}'$ is as \eqref{eq:mrdpf-Q} for the codebook of all $N'$ codewords. Hence, the analysis in Subsection~\ref{subsec:lemma-proof}, with $N'$ in place of $N$ and $\|\mathsf{P} - \mathsf{Q}'\|_{\mathrm{TV}} \leq 2\epsilon$ in place of $T \leq \epsilon$, gives, w.h.p., $\EX_{\mathsf{P}} d_i(S^k_i,\hat{S}^k_i) \leq D_i + \delta_1$ and $\phi_i(p_{S_i}, \mathsf{P}_{\hat{S}_{i,j}}) \leq P_i + \delta_2$ for all $i \in [m]$ and $j \in [k]$.
 
Equivocation analysis is as follows. Fix $A \in 2_*^{[m]}$, and evaluate all quantities under $\mathsf{Q}$. Given $L = l$, $Z^n \sim \prod_{t} p_{Z|V}(z_t|v_t(l))$ is independent of $(K, \bold{S}^k)$. So we have the following Markov chain $\bold{S}^k_A \to L \to Z^n$ and we have:
\begin{align}
&I(\bold{S}^k_A;Z^n) \leq I(L;\bold{S}^k_A), \label{eq:trivial-eqv-bound} \\
&I(\bold{S}^k_A;Z^n) = I(L;Z^n) - H(L) + I(K,L;\bold{S}^k_A) - I(\bold{S}^k_A;K|L) \nonumber \\
&\quad + H(L|\bold{S}^k_A,Z^n). \label{eq:leak-decomp}
\end{align}
We how bound each term (w.h.p) in \eqref{eq:leak-decomp} (see details in \cite[Appendix~A]{kozlov2026itw}).
We denote $\delta_1, \delta_2, ... \geq 0$ terms that vanish as $k \to \infty$ and $\delta, \epsilon \to 0$.
\begin{enumerate}
\item $I(L;Z^n) \leq nI(V;Z) + k\delta_1$;
\item $H(L) = kR$, because $L$ is uniform under \eqref{eq:inner-Q};
\item $I(K,L;\bold{S}^k_A) \leq kI(\bold{S}_A;\hat{\bold{S}}) + k\delta_3$;
\item $I(\bold{S}^k_A;K|L) \geq k\min\{R_k, I(\bold{S}_A;\hat{\bold{S}})\} - k\delta_4$;
\item $H(L|\bold{S}^k_A,Z^n) \leq k[R - \rho_A]^+ + k\delta_5$, where $\rho_A \doteq [I(\bold{S}_A;\hat{\bold{S}}) - R_k]^+ + \frac{n}{k} I(V;Z)$.
\end{enumerate}
By the chain rule and the fact that $x - \min\{R_k, x\} = [x - R_k]^+$, terms 3-4 give $I(L;\bold{S}^k_A) \leq k[I(\bold{S}_A;\hat{\bold{S}}) - R_k]^+ + k\delta_6$. Substituting these bounds into \eqref{eq:leak-decomp} gives $\frac{1}{k} I(\bold{S}^k_A;Z^n) \leq \rho_A - R + [R - \rho_A]^+ + \delta_7 = [\rho_A - R]^+ + \delta_7$, and, with \eqref{eq:trivial-eqv-bound}, we obtain, w.h.p.,
\begin{equation}
\tfrac{1}{k} I(\bold{S}^k_A;Z^n) \leq \min\bigl\{ [\rho_A - R]^+, [I(\bold{S}_A;\hat{\bold{S}}) - R_k]^+ \bigr\} + \delta_6, \label{eq:leak-min}
\end{equation}
i.e., the leakage vanishes if $R \geq \rho_A$, and is at most that of revealing $L$ otherwise.
 
It remains to express \eqref{eq:leak-min} with $c_V$. If $I(V;Y) \geq I(V;Z)$, then $c_V = I(V;Y) - I(V;Z)$, and, since $R = r'I(V;Y) - \delta$, $n \leq r'k + 1$, and $r' > r$, we have $\rho_A - R \leq [I(\bold{S}_A;\hat{\bold{S}}) - R_k]^+ - rc_V + \delta_7$, so the first term in \eqref{eq:leak-min} is at most $[I(\bold{S}_A;\hat{\bold{S}}) - R_k - rc_V]^+ + \delta_7$, as $rc_V \geq 0$. Otherwise, $c_V = 0$, and the second term is $[I(\bold{S}_A;\hat{\bold{S}}) - R_k - rc_V]^+$. In both cases, w.h.p.,
\begin{equation*}
\tfrac{1}{k} I_{\mathsf{Q}}(\bold{S}^k_A;Z^n) \leq [I(\bold{S}_A;\hat{\bold{S}}) - R_k - rc_V]^+ + \delta_8.
\end{equation*}
Finally, as $\|\mathsf{P}_{\bold{S}^k_A Z^n} - \mathsf{Q}_{\bold{S}^k_A Z^n}\|_{\mathrm{TV}} \leq T \leq \epsilon$, the continuity of entropy \cite[Lemma~2.7]{csiszar2011book}, applied to $H(\bold{S}^k_A)$, $H(Z^n)$, and $H(\bold{S}^k_A,Z^n)$, whose alphabet sizes are exponential in $k$, gives $\frac{1}{k} |I_{\mathsf{P}}(\bold{S}^k_A;Z^n) - I_{\mathsf{Q}}(\bold{S}^k_A;Z^n)| \leq \delta_9$. Since $H_{\mathsf{P}}(\bold{S}^k_A) = kH(\bold{S}_A)$, we obtain, w.h.p.,
\begin{align}
\tfrac{1}{k} H_{\mathsf{P}}(\bold{S}^k_A|Z^n) &\geq H(\bold{S}_A) - [I(\bold{S}_A;\hat{\bold{S}}) - R_k - rc_V]^+ - \delta_{10}. \label{eq:inner-eqv-final}
\end{align}
 
The relevant events/bounds analysed above, for all $i \in [m]$ and $A \in 2_*^{[m]}$, are finitely many and each holds w.h.p., so one can fix a codebook for which all of them hold. By \eqref{ineq:inner-secrecy} and \eqref{eq:inner-eqv-final}, and since $n/k \leq r + \epsilon$ and all deltas above vanish as $k \to \infty$ and $\delta, \epsilon \to 0$, \eqref{eq:ach-dp}-\eqref{eq:ach-eqv} hold, i.e., the tuple is achievable.

\newpage

\appendices
\section{}
\subsection{Proofs of the Bounds in the Equivocation Analysis}
All quantities are evaluated under \eqref{eq:inner-Q} for a fixed codebook, and each bound holds w.h.p. over the random codebook.

\emph{Term 1.}
\begin{align}
&I(L;Z^n) \relsm{a}{\leq} I(V^n(L);Z^n) \relsm{b}{\leq} \textstyle\sum_{t=1}^{n} I(V_t(L);Z_t) 
\\ &\relsm{c}{\leq} n\bigl(I(V;Z) + \delta_n \bigr),
\end{align}
where (a) holds since $L \to V^n(L) \to Z^n$; (b) holds since $p_{Z^n|V^n} = \prod_{t=1}^{n} p_{Z|V}$ is memoryless, so $H(Z^n|V^n) = \sum_{t} H(Z_t|V_t)$, while $H(Z^n) \leq \sum_{t} H(Z_t)$, (c) holds asymptotically for $n \to \infty$ with $\delta_n \to 0$.

\emph{Term 3.}
\begin{align}
&I(K,L;\bold{S}^k_A) \relsm{a}{=} I(\hat{\bold{S}}^k(K,L);\bold{S}^k_A) \relsm{b}{\leq} \textstyle\sum_{j=1}^{k} I(\hat{\bold{S}}_j(K,L);\bold{S}_{A,j}) \nonumber \\
&\relsm{c}{\leq} k\bigl(I(\bold{S}_A;\hat{\bold{S}}) + \delta_k \bigr),
\end{align}
where (a) is due to DPI given $(K,L) \to \hat{\bold{S}}^k(K,L) \to \bold{S}_A$; (b) holds since the channel $\prod_{j=1}^{k} p_{\bold{S}_A|\hat{\bold{S}}}$ is memoryless, so $H(\bold{S}^k_A|\hat{\bold{s}}^k(K,L)) = \sum_{j} H(\bold{S}_{A,j}|\hat{\bold{s}}_j(K,L))$, while $H(\bold{S}^k_A) \leq \sum_{j} H(\bold{S}_{A,j})$; (c) holds asymptotically for $k \to \infty$ with $\delta_k \to 0$
 
\emph{Term 4.} Under the idealised distribution \eqref{eq:inner-Q}, $K$ is uniform on $\mathcal{K}$ and independent of $L$, we have:
\begin{equation}
I(\bold{S}^k_A;K|L) = kR_k - H(K|L,\bold{S}^k_A).
\end{equation}
Let $\mathcal{K}_A$ be the list of all $\kappa$ such that $(\hat{\bold{s}}^k(\kappa,L), \bold{S}^k_A) \in \T^k(\hat{\bold{S}},\bold{S}_A)$, let $N_K = |\mathcal{K}_A \setminus \{K\}|$, and let $B = \mathbbm{1}\{K \notin \mathcal{K}_A\}$. Then,
\begin{equation}
H(K|L,\bold{S}^k_A) \leq H(B) + H(K|B,L,\bold{S}^k_A),
\end{equation}
where $H(B) \leq 1$, $H(K|B = 1, L, \bold{S}^k_A) \leq kR_k$, and, given $(L, \bold{S}^k_A)$ and $B = 0$, $K$ lies in $\mathcal{K}_A$, which has size $1 + N_K$. We have
\begin{equation}
H(K|L,\bold{S}^k_A) \leq 1 + \mathsf{Q}(K \notin \mathcal{K}_A) kR_k + \EX_{\mathsf{Q}} \log(1 + N_K). \label{eq:term4-bound}
\end{equation}
Since $\bold{S}^k_A$ is generated from $\hat{\bold{s}}^k(K,L)$ through a memoryless channel, the conditional typicality lemma gives $\EX \mathsf{Q}(K \notin \mathcal{K}_A) \to 0$, so $\mathsf{Q}(K \notin \mathcal{K}_A) \leq \delta/(1 + R_k)$ w.h.p. For every $\kappa \neq K$, the codeword $\hat{\bold{s}}^k(\kappa,L)$ is independent of $\bold{S}^k_A$. Hence, by the joint typicality lemma \cite{elgamal2011book} and the union bound over $\kappa$, $\EX N_K \leq 2^{k(R_k - I(\bold{S}_A;\hat{\bold{S}}) + \delta)}$. By Markov's inequality, $\EX_{\mathsf{Q}} N_K$ is at most $2^{k\delta}$ times this bound w.h.p., and, by Jensen's inequality and $\log(1 + 2^a) \leq [a]^+ + 1$,
\begin{align}
&\EX_{\mathsf{Q}} \log(1 + N_K) \leq \log(1 + \EX_{\mathsf{Q}} N_K) \\
&\leq k[R_k - I(\bold{S}_A;\hat{\bold{S}}) + 2\delta]^+ + 1.
\end{align}
Substituting these bounds into \eqref{eq:term4-bound}, and using $R_k - [R_k - x]^+ = \min\{R_k, x\}$, gives $I(\bold{S}^k_A;K|L) \geq k\min\{R_k, I(\bold{S}_A;\hat{\bold{S}}) - 2\delta\} - k\delta - 2$, which implies term~4, as $\min\{R_k, x - 2\delta\} \geq \min\{R_k, x\} - 2\delta$.
 
\emph{Term 5.} Let $\mathcal{L}_A$ be the list of all $l$ such that $(\hat{\bold{s}}^k(\kappa,l), \bold{S}^k_A) \in \T^k(\hat{\bold{S}},\bold{S}_A)$ for some $\kappa$ and $(v^n(l), Z^n) \in \T^n(V,Z)$, and let $N_A = |\mathcal{L}_A \setminus \{L\}|$. Similarly to \eqref{eq:term4-bound},
\begin{equation}
H(L|\bold{S}^k_A,Z^n) \leq 1 + \mathsf{Q}(L \notin \mathcal{L}_A) kR + \EX_{\mathsf{Q}} \log(1 + N_A).
\label{eq:term5-bound}
\end{equation}
Since $\bold{S}^k_A$ and $Z^n$ are generated from $\hat{\bold{s}}^k(K,L)$ and $v^n(L)$ through memoryless channels, the conditional typicality lemma gives $\EX \mathsf{Q}(L \notin \mathcal{L}_A) \to 0$, so $\mathsf{Q}(L \notin \mathcal{L}_A) \leq \delta/(1 + R)$ w.h.p. For every $l \neq L$, the codewords $\{\hat{\bold{s}}^k(\kappa,l)\}_\kappa$ and $v^n(l)$ are independent of $(\bold{S}^k_A, Z^n)$ and of each other. Hence, by the joint typicality lemma and the union bound over $\kappa$,
\begin{equation}
\EX N_A \leq 2^{kR} \min\{1, 2^{k(R_k - I(\bold{S}_A;\hat{\bold{S}}) + \delta)}\} 2^{-n(I(V;Z) - \delta)},
\end{equation}
Since the minimum is at most $2^{-k([I(\bold{S}_A;\hat{\bold{S}}) - R_k]^+ - \delta)}$ and $n \leq (r' + 1)k$, it follows that $\EX N_A \leq 2^{k(R - \rho_A) + k(2 + r')\delta}$. By Markov's inequality, $\EX_{\mathsf{Q}} N_A \leq 2^{k(R - \rho_A) + k(3 + r')\delta}$ w.h.p., so, by Jensen's inequality, $\EX_{\mathsf{Q}} \log(1 + N_A) \leq k[R - \rho_A]^+ + k(3 + r')\delta + 1$. Substituting these bounds into \eqref{eq:term5-bound} proves the bound for term 5.

\bibliographystyle{IEEEtran}
\bibliography{main}
\end{document}

%% file: model.tex
\tikzset{every node/.style={font=\small}}%
\tikzset{%
blk/.style={draw,rectangle,rounded corners=2pt,inner xsep=2.5pt,inner ysep=1.8pt,minimum height=4.6mm},%
modality/.style={blk},%
coder/.style={blk,minimum height=5.6mm},%
channel/.style={blk,minimum height=5.6mm},%
eve/.style={blk},%
key/.style={blk},%
lbl/.style={draw=none,inner sep=1pt}%
}%
\tikzset{node distance=0}%
\ifx\spc\undefined\newlength{\spc}\fi%
\setlength{\spc}{7mm}%

\providecommand{\vdotsc}{\vbox{\baselineskip3pt\lineskiplimit0pt\hbox{.}\hbox{.}\hbox{.}}}%
\begin{tikzpicture}
\node[modality] (S1) {$p_{S_1}$};
\node[lbl,below=0.3mm of S1] (dots1) {\vdotsc};
\node[modality,below=0.3mm of dots1] (Si) {$p_{S_i}$};
\node[lbl,below=0.3mm of Si] (dots2) {\vdotsc};
\node[modality,below=0.3mm of dots2] (Sm) {$p_{S_m}$};

\node[coder,right=1.5\spc of Sm] (E) {Encoder};
\draw[->] (Si.east) to[out=0,in=90] node[midway, above] {$\{S_i^k\}_{i \in E}$} (E.north);

\node[channel,right=\spc of E] (C) {$p_{YZ|X}$};
\draw[->] (E) -- (C) node[midway, above] {$X^n$};

\node[coder,right=\spc of C] (D) {Decoder};
\draw[->] (C) -- (D) node[midway, above] {$Y^n$};

\node[eve,above=4mm of C] (EV) {Eve};
\draw[->] (C) -- (EV) node[midway, right] {$Z^n$};

\node[key,above=1.8mm of EV] (K) {$K$};
\draw[->] (K.west) to[out=180,in=90] (E.north);
\draw[->] (K.east) to[out=0,in=90] (D.north);

\draw[->] (D.east) -- ++(1.85\spc,0) node[midway, above] {$\{\hat{S}_i^k\}_{i \in [m]}$};

\node[draw=black, dashed, rounded corners=3pt, fit=(S1) (dots1) (Si) (dots2) (Sm), inner sep=1mm, label={[lbl,label distance=0.4mm]above:Modalities}] (modgroup) {};

\node[lbl, below=0pt of C] {Channel};
\node[lbl, above=0pt of K] {Secret Key};
\end{tikzpicture}%